\documentclass[conference,twocolumn,10pt]{IEEEtran}

\usepackage{threeparttable}

\usepackage{amsmath}
\usepackage[noadjust]{cite}
\usepackage{enumerate}
\usepackage{amssymb}
\usepackage{graphicx}
\usepackage{float}
\usepackage{epstopdf}
\usepackage{pbox}
\usepackage{color}
\usepackage{varwidth}
\usepackage{algorithm}
\usepackage{algpseudocode}

\usepackage[section,subsection,subsubsection]{extraplaceins}
\usepackage{multirow}
\usepackage{subfig}
\usepackage{booktabs}
\usepackage{citesort}

\newcommand{\RNum}[1]{\uppercase\expandafter{\romannumeral #1\relax}}

\usepackage{arydshln}
\usepackage{threeparttable}
\usepackage[cmintegrals]{newtxmath}
\usepackage{array}

\begin{document}

\title{{Multi-hop Routing with Proactive Route Refinement for 60~GHz Millimeter-Wave Networks }}

{\author{\IEEEauthorblockN{Chanaka Samarathunga\IEEEauthorrefmark{1}, Mohamed Abouelseoud\IEEEauthorrefmark{2}, Kazuyuki Sakoda\IEEEauthorrefmark{3}, Morteza Hashemi\IEEEauthorrefmark{1}}\IEEEauthorblockA{\IEEEauthorrefmark{1}Department of Electrical Engineering and Computer Science, University of Kansas \\ \IEEEauthorrefmark{2}Sony R\&D Center US, San Jose lab, \IEEEauthorrefmark{3}Sony R\&D Center Japan, Tokyo lab}}}
\maketitle

\begin{abstract}
Fundamental requirements of mmWave systems are peak data rates of multiple Gbps and latencies
of the order of at most a few milliseconds. However, highly directional mmWave links are susceptible to frequent link failures under stress conditions such as mobility and human blockage. Under these conditions, multi-hop routing can achieve reliable and robust performance. In this paper, we consider multi-hop millimeter wave (mmWave) wireless systems and propose proactive route refinement schemes that are particularly important under dynamic scenarios. First, we consider the AODV-type protocols and propose a cross-layer approach that integrates sectorized communication at the MAC layer with on-demand multi-hop routing at the network layer. Next, we consider Backpressure routing protocol, and enhance this protocol with periodic HELLO status messages. 
 System-level simulation results based on the IEEE 802.11ad standard are provided that confirm the  benefits of proactive route refinement for the ADOV-type and Backpressure routing protocols. 
\end{abstract}

\IEEEpeerreviewmaketitle


\section{Introduction}

The growing demands for applications with extremely high
data rates as well as the increasing density of wireless devices are catalyzing a coming spectrum crisis in the
sub-6 GHz bands. The spectrum-rich millimeter-wave (mmWave) frequencies between 30 GHz to 300 GHz
have the potential to alleviate the spectrum crunch that the wireless and cellular operators are already experiencing \cite{rappaport2013millimeter,boccardi2013five}. Indeed, this major potential of the mmWave bands has made them the most important component of future mobile cellular and emerging WiFi networks with Gbps data rates. 

Compared with the legacy wireless systems operating in the sub-6 GHz bands, there are significant
challenges that need to be overcome before practical mmWave systems can be commercialized. Propagation
loss at mmWave frequencies is much higher due to a variety of factors including atmospheric absorption and
low penetration. 
In addition to  large path losses,  due to small wavelengths in the mmWave band, most objects such as human body can significantly attenuate the mmWave signals (up to 20 dB), which can entirely break the link. In order to mitigate the blockage issue, there have been several proposals on exploiting reflection paths from walls \cite{genc2010robust}, using intelligent reflecting surfaces \cite{qingqing2019towards}, and  integrating mmWave with lower frequencies \cite{yao2019integrating,hashemi2018out}. One effective approach to combat blockage in mmWave is to leverage multi-hop routing. {In \cite{samarathunga2020benefits}, the authors propose a hop-by-hop multi-path routing protocol that is efficient and fast in switching to a reserved ready-to-use path towards the destination.} However, due to the dynamic conditions in mmWave propagation environment, it is highly likely that the blockages are temporary and highly dynamic. For example, the frequency and duration of  blockage can be characterized using a Poisson Point Process (PPP) technique  \cite{jain2018millimeter}. Thus, \emph{it is desirable that multi-hop routes towards the destination be refined within a shorter period of timescale compared with the routing table reset timescale across the network.} In this paper, we propose proactive route refinement schemes for multi-hop mmWave networks by considering two general classes of multi-hop routing protocols: (i) \emph{AODV-type protocols} that are based on distributing route request (RREQ) and route reply (RREP) messages, and (ii) \emph{Backpressure-type protocols} that are built upon finding the best route based on local information at each node \cite{moeller2010routing}.  





In order to achieve proactive route refinement for the AODV-type protocols, we pose the following question: \emph{given that the sector sweep operation is needed for establishing and maintaining directional mmWave links, is it possible to leverage sector sweep (SSW) frames for routing purposes?} By carefully embedding and integrating the route request and route reply messages with  sector sweep frames, we achieve \emph{a proactive route refinement step} that  can be added to on-demand AODV-type routing protocols.  Leveraging sector sweep frames results in more optimized routes without sending more control messages. However, it should be noted that each SSW frame will be larger since route refinement fields  piggyback on the SSW frames.  




In the backpressure-type protocols, there is no explicit route request or route reply messages (as described in Section \ref{sec:backpressure}). Instead, each node selects one of its neighbor with the maximum backpressure weight and forwards packets to that neighbor. In omni-directional wireless systems,  nodes can overhear other nodes' transmission and extract relevant information that are needed for calculating the backpressure weight. However, this mechanism does not work in mmWave networks due to directionality. Therefore, we propose integrating periodic HELLO messages into backpressure to distribute the parameters that are needed for calculating backpressure weight. This periodic status update message provides the possibility of refining multi-hop routes based on fresh information from neighbor nodes.

To demonstrate the operation and benefits of our proposed route refinement mechanisms, we implement the multi-hop protocol along with the associated route refinement features. The simulation setup is mainly focused on indoor 60 GHz systems where 7 GHz unlicensed band is available. This
 is aligned with nascent industry efforts such as the IEEE 802.11ay standardization activities.  In summary the main contributions of this work are as follows: (i) we propose a cross-layer route refinement mechanism for AODV-type protocols, (ii) we propose integrating a periodic status message into Backpressure protocols, and (iii) we implement the AODV and Backpressure protocols on top of the IEEE 802.11ad standard to demonstrate the benefits of route refinement. 

\section{Background and Related Work}
\label{sec:related-work}
\textbf{Directional Routing Overview:} In general, AODV-type protocols use route  request  (RREQ) and route reply (RREP) messages to establish multi-hop routes between the source STA and the destination STA. RREQ messages are generated by the source and sent to the neighbor STAs. In response, route reply (RREP) messages are generated by the destination STA, and relayed by the intermediate STAs until it is received by the source STA.  
There have been several works to customize the AODV protocol for directional communications. The authors in \cite{choudhury2003impact} evaluate the performance of dynamic source routing (DSR) protocol when executed over directional antennas. 
The simulation results show that routes with a fewer number of hops can be found. On the other hand, neighbor discovery becomes more complex since it may require the antenna system to sweep its transmitting beam sequentially over multiple directions. The work in \cite{gossain2006drp} proposes a Directional Routing Protocol (DRP) that couples some aspects of the routing layer with the MAC layer. 
The authors in \cite{shen2013routing} provide a comparative view 
for several directional routing protocols, including DRP, Directional Dynamic Source Routing (DDSR), Directional Ad-hoc On-demand Distance Vector (DAODV), Energy Efficient Directional Routing (EEDR), and Directional Antenna Multipath Location Aided Routing (DA-MLAR). Moreover, \cite{bandyopadhyay2001adaptive} proposes an adaptive MAC protocol, where each node keeps certain neighborhood information dynamically through the maintenance of an Angle-SINR table, which can improve the performance of directional routing protocols. Another directional routing protocol is proposed in \cite{liu2007directional} that is based on the angle of arrival estimation such that the routing paths are chosen to minimize interference in the network.   

\begin{figure}[t]
	\centering
	\vspace{-.3cm}
	{\includegraphics[scale=.7]{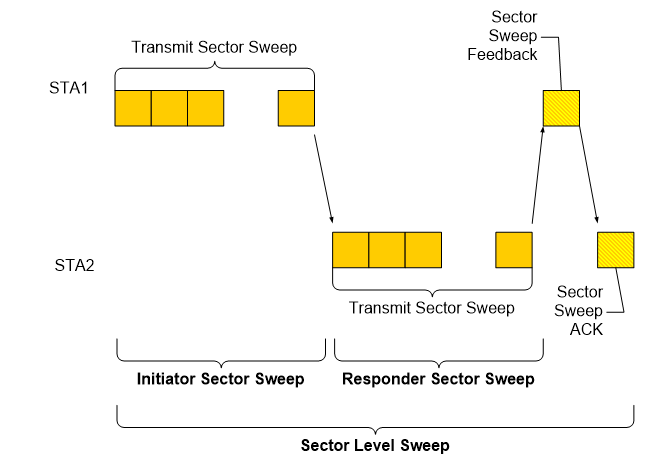}}
	\caption{Sector sweep operation between two STAs to learn the best direction for transmission and receiving mmWave signals.}
	\label{fig:ssw}
\end{figure}

\textbf{Sector Sweep Overview:} To alleviate large path losses in mmWave, large antenna arrays with much smaller form factors can focus the signal energy toward a
specific direction and create directional links. The transmitter and receiver perform sector sweep operation to learn about the best direction and achieve beam alignment. According to the IEEE 802.11ad and 11ay standards, during the sector level sweep (SLS), a pair of STAs exchange a series of sector sweep (SSW) frames (or beacons) over different antenna sectors to find the one providing highest signal quality. The station that transmits first is called the initiator, the second is the responder. The exchange of SSW frames is shown in Figure \ref{fig:ssw} where STA1 is the initiator and STA2 is the responder.  Each STA transmits a series of sector sweep (TXSS) signals to let neighbor STAs to learn received signal quality of each antenna sector. 
During a TXSS, SSW frames are transmitted on different sectors while the pairing node (the responder) receives with a quasi-omni directional pattern. Neighbor STAs listen to the TXSS signals, and report back with SSW feedback to let their neighbor STAs know signal quality of each antenna sector. 
The responder determines the antenna array sector from the initiator that provided the best SNR. 

\textbf{Our Contributions:} Complementing the previous works on directional multi-hop routing, this paper is aimed to answer this question: \emph{how can we achieve proactive route refinement for on-demand routing protocols?} Route refinement is essential for on-demand protocols since mmWave link blockage can be temporary and highly dynamic. For instance, the authors in \cite{jain2018millimeter} have shown that the LOS blockage on average lasts for about $1/ \mu$ seconds, where $\mu =2$ is used in their numerical results \cite{jain2018millimeter}.  By deploying an on-demand routing protocol, the source node has already established a route toward the destination via the relay STA. Thus, the source node would not search for a {better route} (in terms of a pre-defined route metric) until the next global routing table reset. Therefore, it is desirable to provide agile route refinement solutions under dynamic blockage scenarios. For the AODV-type protocol,  we propose to leverage the SSW frames that are being sent according to a transmit schedule. The enhanced SSW frames carry routing-related fields and elements. For the Backpressure-type protocol, we integrate a periodic HELLO message that is sent to the neighbor nodes.

\vspace{-.2cm}
\section{Cross-Layer Route Refinement for AODV}
In order to quickly establish a multi-hop route toward the destination, the originating STA sends route request (RREQ) to its neighbor STAs, assuming that the STAs have performed the SSW beforehand, and that there are periodic sector sweep operation for link maintenance purposes.
Once the blockage occurs, the source node sends route request frames toward its neighbors (i.e., the relay node). In this case, the generated RREQ message is passed to the MAC layer of the source STA for delivering to the relay STA. By deploying an on-demand routing protocol, the RREQ frame is delivered to the relay node as soon as possible in order to restore communication between the source and destination STAs. Therefore, transmitting directional RREQ messages and receiving directional RREP messages enable the source STA to find an alternative route toward the destination.

\begin{algorithm}[t]
\caption{Enhanced-SSW -- Initiator}
\begin{varwidth}{\dimexpr\linewidth-2\fboxsep-2\fboxrule\relax}
\begin{algorithmic}[1]
\State $\mathcal{S}$: set of neighbor nodes
\Function{Transmit Sector Sweep}{}
\If {RREQ is received from the Network layer}
\State Add RREQ fields to legacy SSW frame 
\State Create an {Enhanced-SSW} frame 
\State Set \texttt{transmitFrame =} Enhanced-SSW frame   
\Else
\State \texttt{transmitFrame = } Legacy-SSW frame
\EndIf
\For {neighbor node $i \in \mathcal{S}$}
\State Send \texttt{transmitFrame} to node $i$
\EndFor
\EndFunction
\State 
\Function{Receive Sector Sweep Feedback}{}
\State Process the frame to update sectors information
\If {Enhanced-SSW feedback frame is received}
\State Extract routing fields and update the routing table
\EndIf
\EndFunction
\end{algorithmic}
\end{varwidth}%
\label{alg:source-SSW}
\end{algorithm} 

\begin{algorithm}[t]
\caption{Enhanced-SSW -- Respondor}
\begin{varwidth}{\dimexpr\linewidth-2\fboxsep-2\fboxrule\relax}
\begin{algorithmic}[1]
\Function{Receive Sector Sweep}{}
\State Receive the transmitted sector sweep frames 
\If {Enhanced-SSW frame is received}
\State Extract routing fields 
\State Query local routing table 
\State Embed routing information in an Enhanced sector sweep feedback frame 
\Else 
\State Reply with legacy sector sweep feedback frame.
\EndIf
\EndFunction
\end{algorithmic}
\end{varwidth}%
\label{alg:relay-SSW}
\end{algorithm} 

\begin{figure}[t]
	\centering
	{\includegraphics[scale=.38, trim=0cm 0.2cm .2cm 0.0cm, clip]{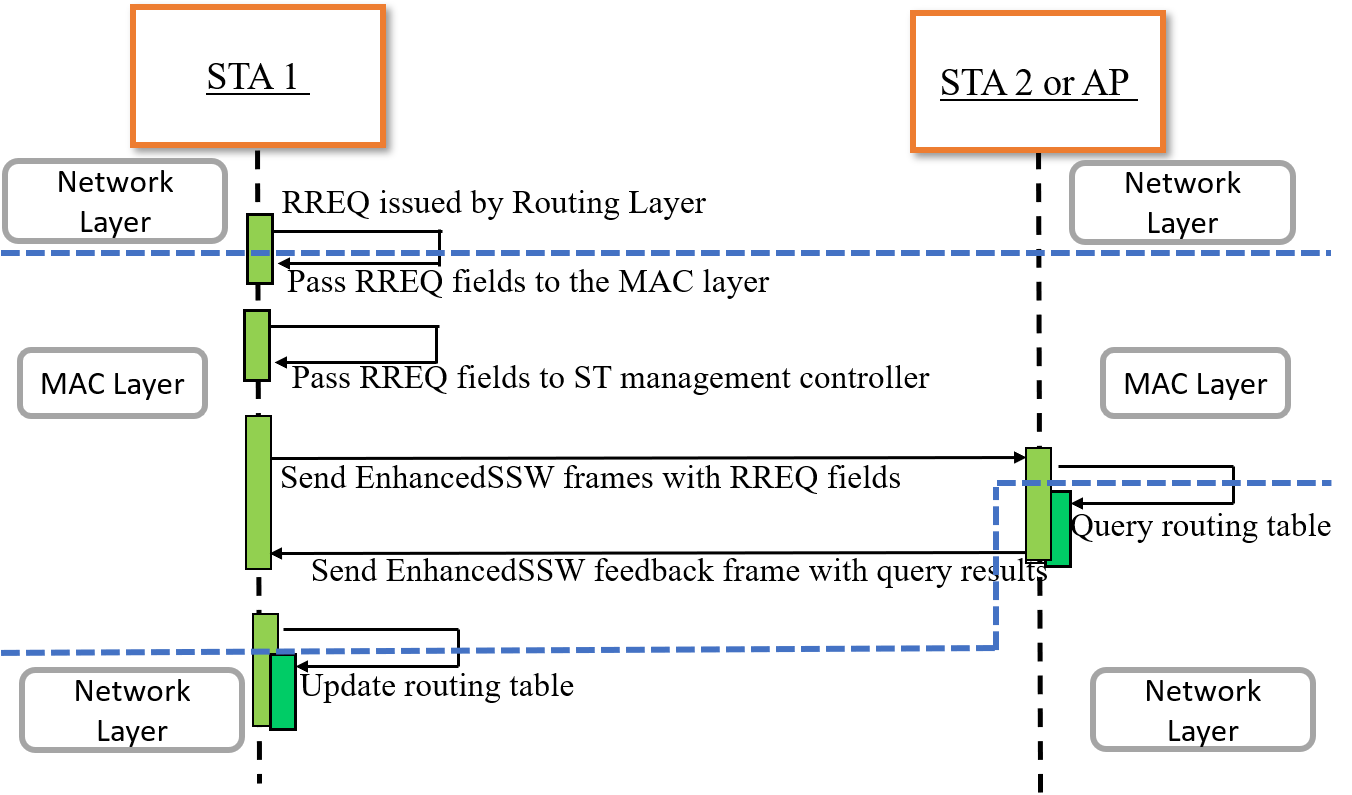}}
	\caption{Message sequence diagram for integrating  the SSW frames and route refinement messages.}
	\label{fig:sequence}
\end{figure}

 In order to utilize SSW frames for route refinement, we propose the following protocol: by deploying an on-demand routing protocol, the source STA takes three steps: (1) it sends the RREQ frame toward the relay node (i.e., the normal operation of the on-demand routing), (2) it extracts and stores the RREQ elements, and (3) it will use the RREQ elements at the next SSW transmission opportunity.
 At the next SSW interval, the source STA adds the RREQ elements to the SSW frames. We refer to this type of the SSW frame as \emph{Enhanced-SSW} frame. The responder node receives the Enhanced-SSW frames across different sectors, extracts the route discovery elements, and queries its routing table for finding a potential route toward the destination requested through the Enhanced-SSW frames. The responder node  replies to the transmitter node with the Enhanced-SSW reply frames, which potentially includes route information toward the destination. Enhanced-SSW frames are received by the source node that extracts route information and updates its routing table.   Note that this sector sweep operation is performed with all neighbor nodes. As a result, sector sweep frames -- which are used for establishing/maintaining directional links -- can potentially lead to refined and optimized multi-hop routes toward the destination. Algorithms \ref{alg:source-SSW} and \ref{alg:relay-SSW} summarize the steps at the initiator and respondor STAs.



 
Figure \ref{fig:sequence} shows the sequence diagram of the implementation, where RREQ messages are passed to the MAC layer, embedded in the SSW frames, and received by a neighbor STA or access point (AP). In response, the neighbor STA or AP queries its own local routing table, and embeds the query results in the reply SSW frames. Upon receiving the reply SSW, the initiator STA  updates its routing table. We see that there are several rounds of cross-layer message passing at the initiator and responder nodes. From these steps, we note that route refinement messages piggyback on the SSW frames, and thus they do not introduce overheads in terms of the number of messages. The overhead of transmission and reception of routing control messages is translated to function calls between the MAC layer and routing layer at the STA nodes, i.e., query and update the routing table. This method achieves a cross-layer route refinement for directional communication. However, it should be noted that adding RREQ/RREP increases the size of legacy SSW frames. 

\textbf{Discussion:} Beam refinement and tracking are also utilized under dynamic and mobile scenarios for link maintenance. In this paper, we only consider leveraging the SSW frames for route refinement, and beam refinement integration with routing operation is out of scope of this work. Moreover, we note that the route refinement is performed at the link layer and on a per-link basis meaning that Enhanced-SSW frames can be exchanged between the source and relay node, two relay nodes,  or relay and destination node. Therefore, the proposed scheme is not limited to the last hop only.  If the responder STA does not have more-optimized route information to send to the initiator, then Enhanced-SSW frame exchange would not modify the routing table at the initiator STA. 

\begin{algorithm}[t]
\caption{mmWave-BCP}
\begin{varwidth}{\dimexpr\linewidth-2\fboxsep-2\fboxrule\relax}
\begin{algorithmic}[1]
\State $\mathcal{S}_i$: Set of neighbor nodes for node $i$
\For {$j \in \mathcal{S}_i$}
\State {Receive periodic HELLO message from node $j$} 
\State Compute backpressure weight $w_{i,j}$ for neighbor $j$
\State Find the neighbor $j^*$ such that $j^*=\operatorname*{arg\,max}_j w_{i,j}$
\If {$w_{i,j*}$ $>0$}
\State Transmit packets to $j^*$ until the next HELLO interval
\Else
\State Wait for a reroute period and go to line 3
\EndIf
\EndFor
\end{algorithmic}
\end{varwidth}
\label{alg:BCP}
\end{algorithm} 
\section{Route Refinement for Backpressure Protocol}
\label{sec:backpressure}
A backpressure routing algorithm is based on solving a problem known as MaxWeight, where the goal is to maximize the weighted sum of link rates. The weights are defined by backlog differentials between neighbor nodes.
Backpressure algorithm leads to the problem of minimizing the Lyapunov drift that is defined as the difference between the values of the Lyapunov function at the current time slot and at the next time slot. In order to solve the MaxWeight problem, intuitively data packets are sent over links with high rates and to neighbors with small queue lengths. For instance, Backpressure Collection Protocol (BCP) \cite{moeller2010routing} is one version of the backpressure algorithm that can be implemented in a distributed manner. Then, each node independently makes routing decisions based on local information. Let $\mathcal{Q}_i$ represent the queue length at node $i$. 
Then $\Delta\mathcal{Q}_{i,j}=\mathcal{Q}_i -\mathcal{Q}_j$ is the queue differential (backpressure) between node $i$ and its neighbor node $j$. Let  $\overline{\mathcal{R}}_{i\rightarrow j}$ denote the estimated link rate from $i$ to $j$ and $\overline{ETX}_{i\rightarrow j}$ be the average number of transmissions for a packet to be successfully sent over the link. In the routing policy of BCP, node $i$ calculates the following backpressure weight for each neighbor $j$:
\begin{equation}
 w_{i,j} = (\Delta \mathcal{Q}_{i,j} - V\cdot \overline{ETX}_{i\rightarrow j}) \cdot \overline{\mathcal{R}}_{i\rightarrow j},   
\end{equation}
{where $V > 0 $ is a non-negative control parameter to adjust the importance of the penalty function $\overline{ETX}_{i\rightarrow j}$.}
The routing decision (next hop of the packet) is determined by finding the neighbor $j^*$ with the highest weight. Then the node needs to make the forwarding decision: if $w_{i,j^*}>0$, the packets are forwarded to node $j^*$.

In omni-directional systems, in order to disseminate all the necessary information to compute backpressure weights, BCP header fields include local queue information that are broadcasted. Therefore, all nodes within reception range of the transmitter receive and process the BCP packet header through the snoop interface. This method, however, does not work for mmWave systems due to directionality of transmission and reception. In addition, in contrast to the AODV protocol, there is no RREQ/RREP messages in BCP. In other words, BCP routing is achieved based on local information of each node from its neighbor nodes.    In order to mitigate these issues and achieve proactive route refinement for the BCP protocol, we propsoe to add a periodic HELLO message to the original BCP protocol. Thus,  all the necessary information to compute weights is exchanged amongst nodes by means of periodic emission of HELLO messages. It should be noted that under dynamic scenarios where nodes are joining or leaving the network or blockage occurs frequently, the HELLO interval needs to be set to a small value.  On the other hand, a larger HELLO interval will be sufficient for more stationary network conditions. In Section \ref{sec:simulation}, we examine different values of HELLO interval and its effect on the system performance. 
Algorithm \ref{alg:BCP} summarizes the mmWave-BCP protocol.

 \begin{figure}[t]
 	\centering
 	\subfloat[Indoor mmWave setup]{
 		\includegraphics[scale=.47, trim = 2cm 2cm 2cm 1.5cm, clip]{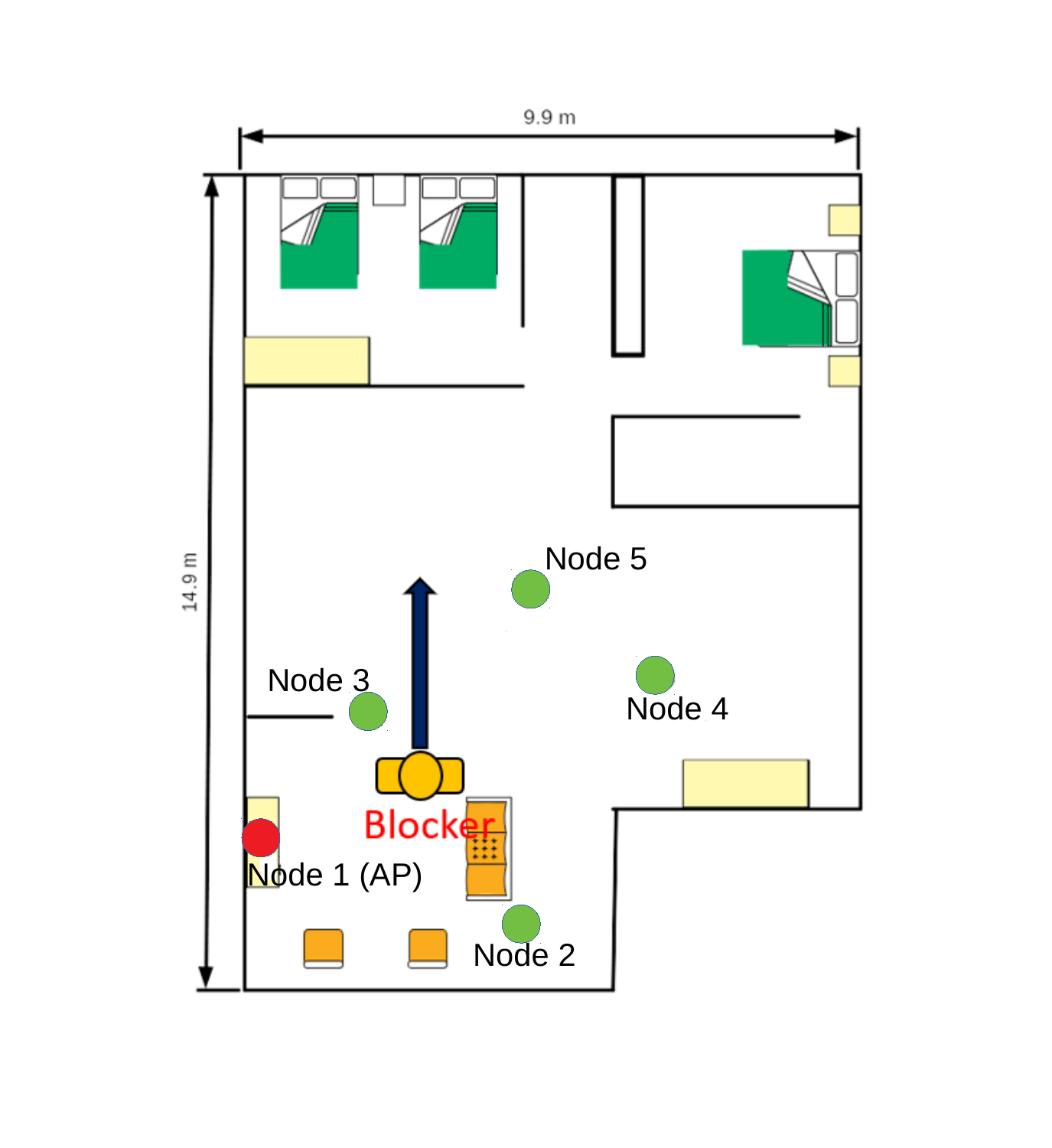}
 		\label{fig:room}
 	} 
 	\hspace{.25cm}
 	\subfloat[Network topology and data flows]{\includegraphics[scale=.48, trim = 1.5cm 2.1cm 1cm 1.5cm, clip]{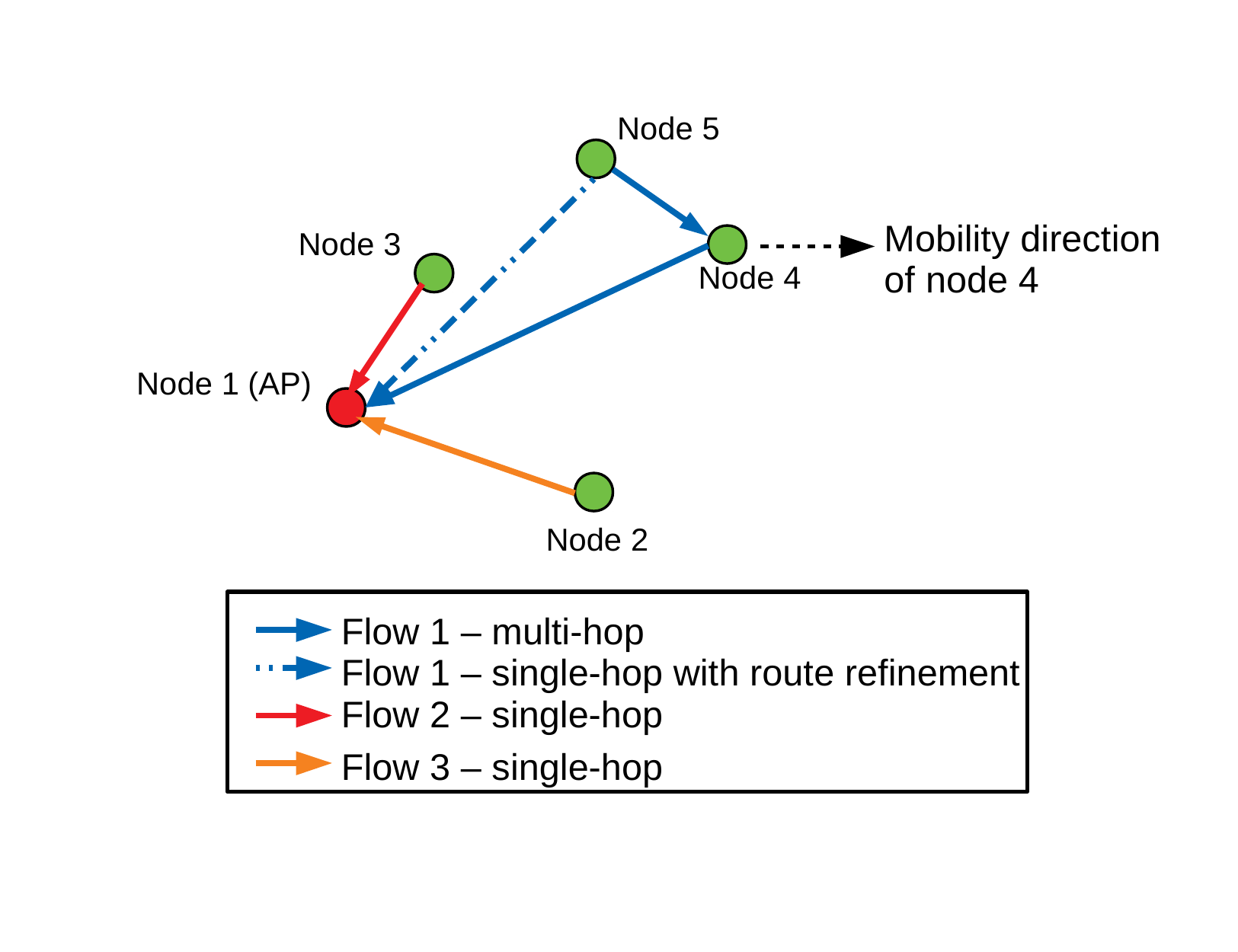}
 		\label{fig:room-raytracing}
 	}
 	\caption[=]{\small{Indoor mmWave network with a human blocker and mobility.}}
 \end{figure}
 \vspace{-.2cm}
\section{Simulation Results}
\label{sec:simulation}
\begin{figure*}[t]
 	\centering
 	 	\vspace{-.5cm}
 	\subfloat[\emph{Throughput performance} ]{  \includegraphics[scale=.3]{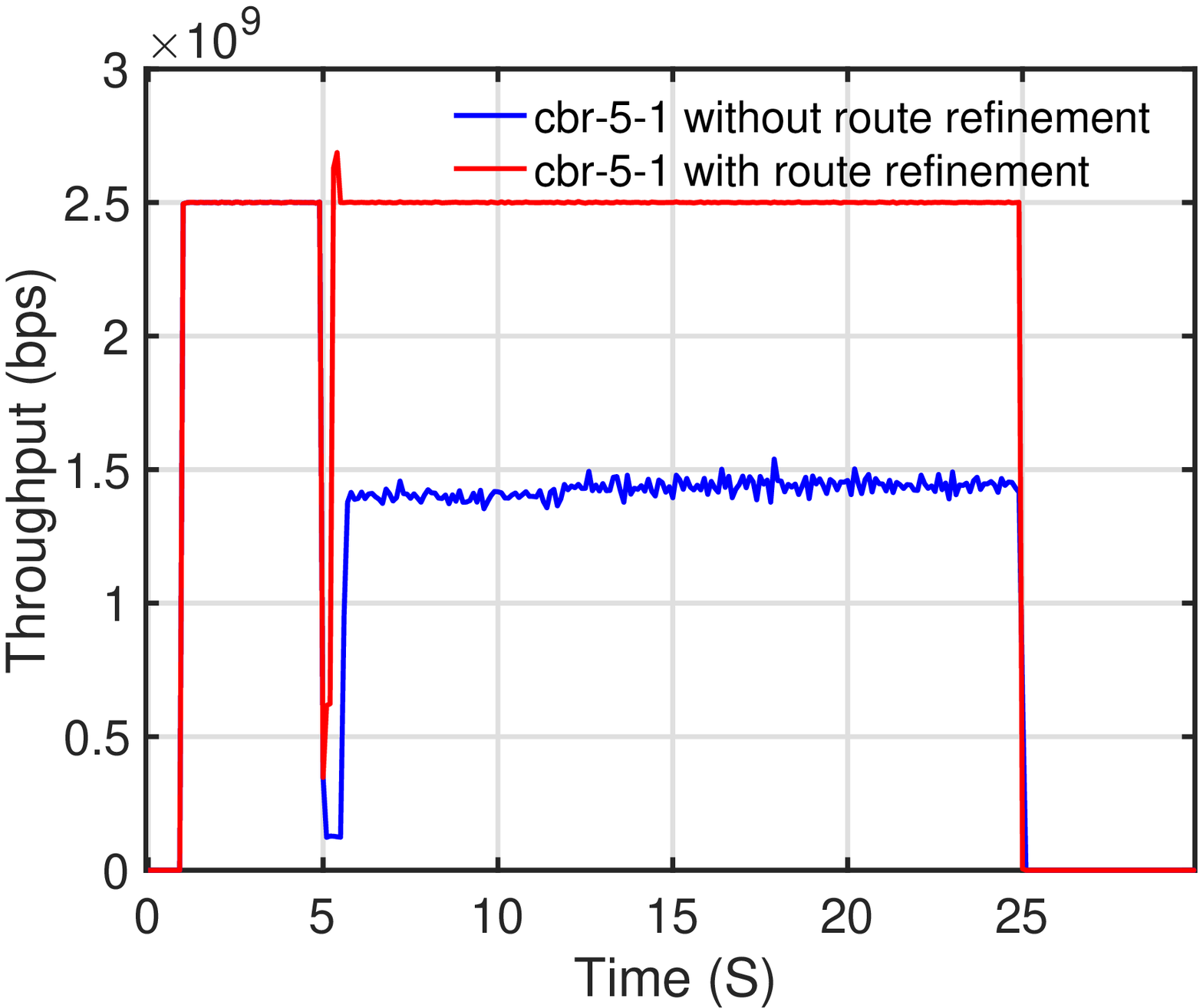}
 		\label{fig:C612-Tput}
 	}
 	\subfloat[\emph{Delay performance}]{
 		\includegraphics[scale=.3]{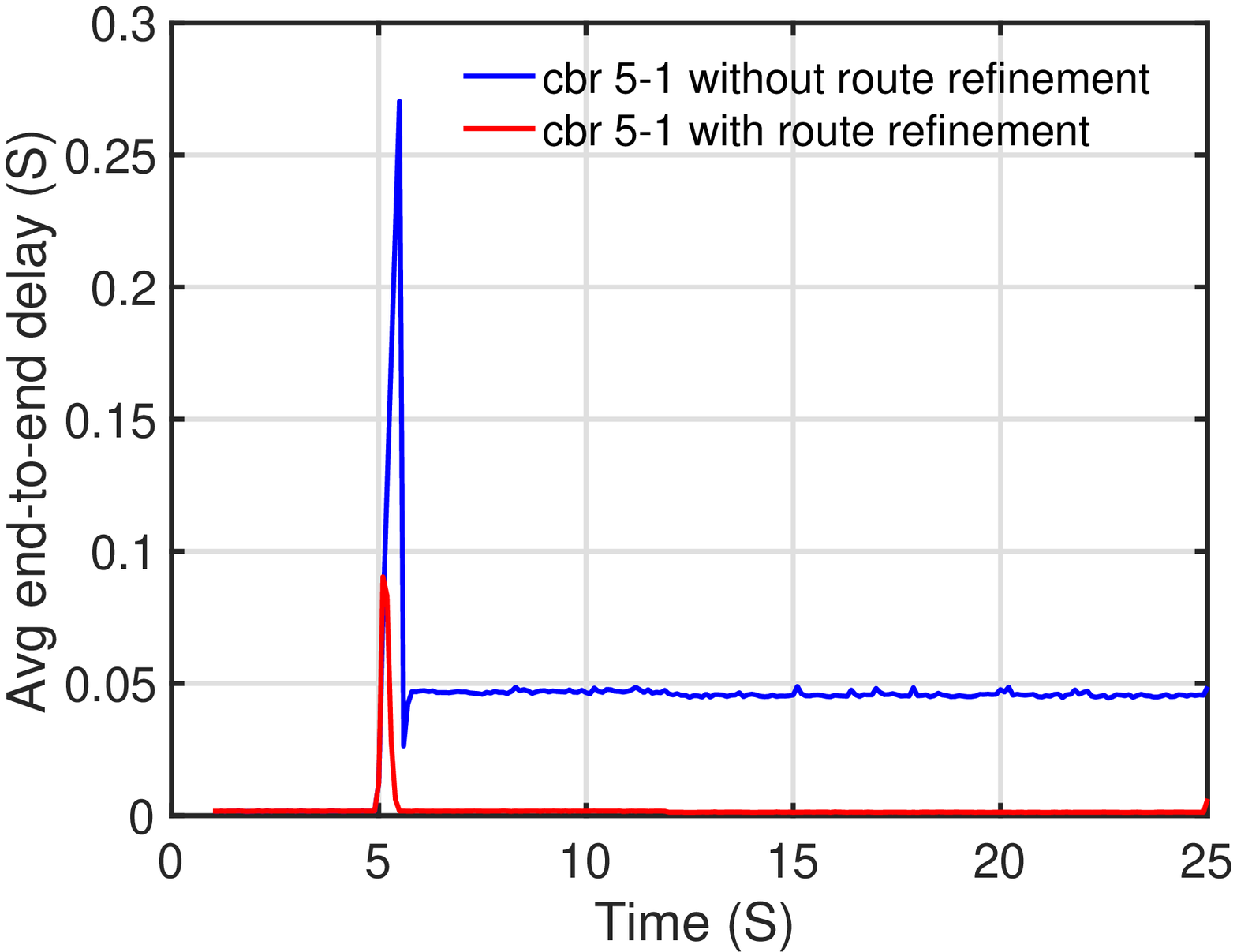}
 		\label{fig:C612-delay}
 	} 
 	\subfloat[\emph{CDF of delay}]{\includegraphics[scale=.3]{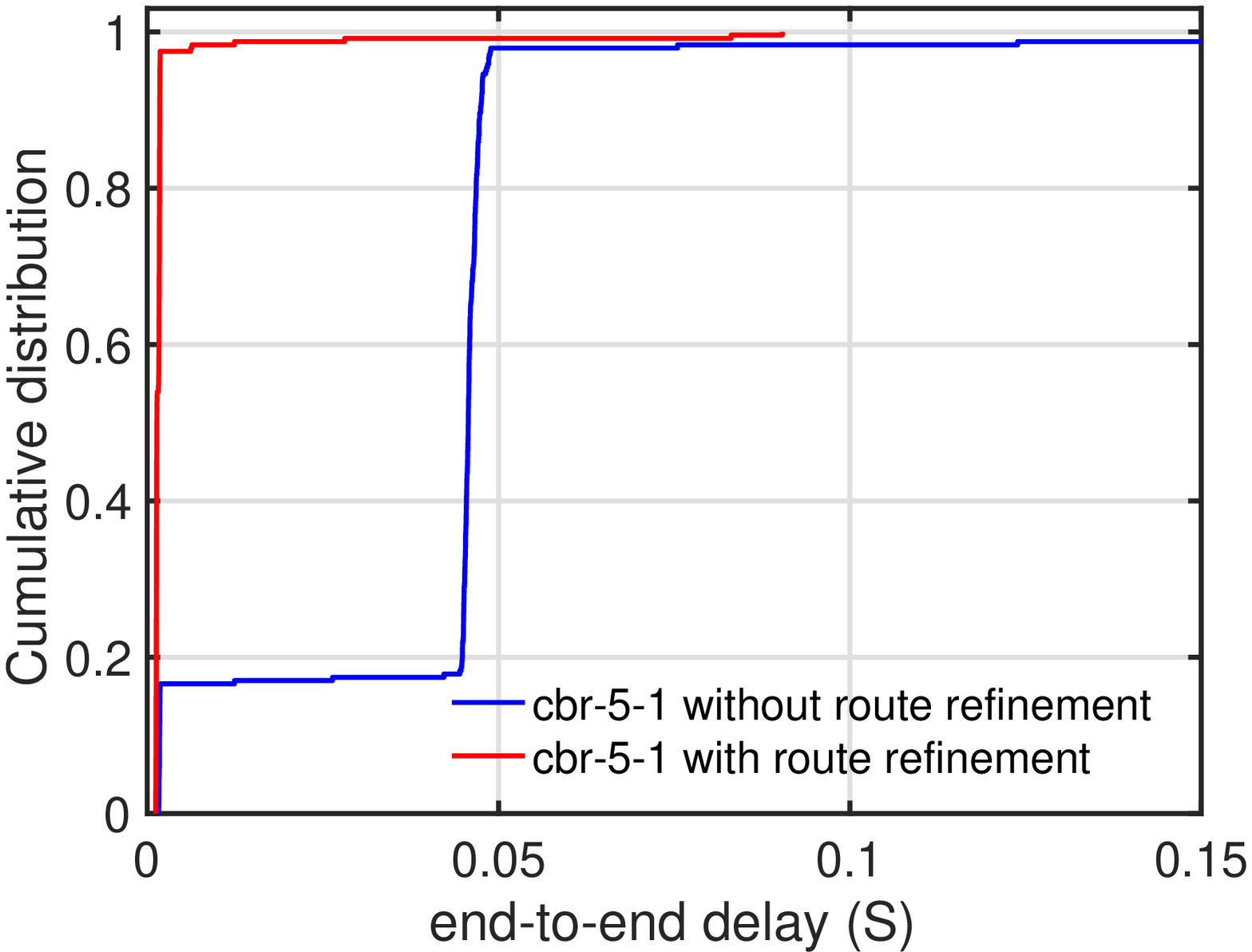}
 		\label{fig:C612-cdf-delay}
 	}
\vspace{-.2cm}
 	\caption[=]{\small{Performance of \textbf{AODV routing protocol} with and without route refinement using sector sweep frames. Blockage lasts for 200ms. }}
 \end{figure*}
 In order to demonstrate the benefits of on-demand routing with proactive route refinement, we consider a mmWave network as shown in Figure \ref{fig:room} that includes human blockage and node mobility. 
 Nodes 1, 2, 3 and 5 are stationary, and node 4 is mobile with the mobility pattern shown in Figure \ref{fig:room-raytracing}. 
 All nodes are equipped with the IEEE 802.11ad MAC and SC PHY specifications with AWGN channel model. It should be noted that although we are performing system-level simulations on top of the IEEE 802.11ad, the underlying assumption is that all nodes in the network are capable of sending and receiving Enhanced-SSW frames. 
 The simulation parameters are summarized in Table \ref{tab:parameter}. 
In order to model the propagation environment, we use Remcom X3D ray tracer with High Fidelity Propagation Model (HFPM) enabled. The total number of computed paths is set to 25 with the number of reflections equal to 3, number of diffraction 1, and number of transmissions 3. Materials used for simulating the building along with their properties are listed in Table \ref{tab:material}.

\begin{table}[t]
	\caption{\small{Simulation parameters}}
 	\centering
 	\vspace{-.2cm}
  	\begin{tabular}{@{}lc@{}}\toprule
  	\textbf{Simulation Parameter} & \textbf{Value} \\ \midrule
  	Transmit power & 18dBm  \\
      Preamble detection threshold & -68dBm \\
      Noise level & -70.6dBm  \\
      Energy detection threshold & -48dBm  \\
      Channel access scheme & Contention based \\
      Beacon interval (BI) & 100ms  \\
      Beacon header interval (BHI) & 5ms  \\
      Data transmission interval (DTI) & 95ms  \\
      Rate controller & ARF \\
      Maximum number of aggregated MPDU & 64 \\
      Transmit opportunity duration (TXOP) & 300 $\mu$s \\
      Human blocker path loss & 20dB \\
      Human blocker dimensions (length, width, height) & (0.5m, 0.5m, 1.8m) \\
 	\bottomrule
 	\end{tabular}
 	\label{tab:parameter}
 \end{table}
\begin{table}[t]
 	\caption{\small{Materials characteristics  used in the simulations}}
\begin{threeparttable}
 	\centering
 		\vspace{-.2cm}
  	\begin{tabular}{@{}lccc@{}}\toprule
  	\textbf{Material} &  \textbf{Relative Permittivity\tnote{1}} & \textbf{Conductivity\tnote{2}} & \textbf{Thickness~(m)} \\ \midrule
 Brick wall & 5.31 & 0.8967 & 0.3\\
  Concrete wall & 5.31 & 0.8967 & 0.3 \\
  Wood & 1.99 & 0.3784 & 0.03\\
  Glass & 6.27 & 0.5674 & 0.001\\
	\bottomrule
 	\end{tabular}
 \begin{tablenotes}
       \item[1]Relative permittivity with respect to free space or vacuum.
       \item[2]Conductivity is measured in terms of Siemens per meter (S/m). 
     \end{tablenotes}
 \end{threeparttable}
 	\label{tab:material}
 \end{table}

\subsection{Blocker Scenario with Node Mobility and Single Data Flow}
\label{sec:single}
First, we activate nodes 1, 4 and 5, while nodes 2 and 3 are not active. Human blockage is modeled as an additional $20$ dB path loss that is applied at time $5$s. To model dynamic changes in the environment, the link between the source and destination (sink) node is blocked for $200$ms, and after that the blockage is removed.  The generated traffic data rate at node 5 is 2.5 Gbps with a constant bit rate (CBR) pattern. Figure \ref{fig:C612-Tput} depicts the throughput performance measured at node 1. Before time 5 seconds, node 5 transmits directly to node 1 and the achieved throughput is 2.5 Gbps. Once the blockage happens, the AODV routing protocol kicks in to find an alternative route toward the destination node, in which case the data traffic is routed to node 4 as a relay node. From the results, we see that if there is no route refinement, the throughput remains the same even after the blockage is removed.  On the other hand, once the blockage is removed at 5.2 second and by activating the proactive route refinement, node 5 can switch back to a single-hop route by directly transmitting to node 1 and achieving the 2.5 Gbps throughput. Note that this route refinement step is achieved by the sector sweep operation between node 5 and node 1 once the blockage has been removed. Therefore, no additional signal exchanges is needed. Figure \ref{fig:C612-delay} and \ref{fig:C612-cdf-delay} compare the delay and CDF of delay for multi-hop routing with and without route refinement. From the results, we observe that route refinement significantly improves the delay performance.

Next, we deploy the mmWave-BCP routing protocol and investigate route refinement using HELLO packets in order to distribute up-to-date queue length and data rate information for calculating the backpressure weights. In this simulation, we increase the blockage duration to 1 second (i.e., between 5 to 6 seconds). The results shown in Figure \ref{fig:hello-bcp} depicts the throughput, delay, and delay CDF for two HELLO intervals of 1 and 5 seconds. From the results, we observe that HELLO messages play an important role to find and switch back to a single-hop topology when the blockage is removed. There is a trade-off between the HELLO message overheads and route refinement agility. 
{\subsection{Blocker Scenario with Multiple Data Flows}
\label{sec:multiple}
Next, the data flows from node 3 and node 2 to node 1 (AP) are activated. The data rates are reduced to be within the capacity region, and are set to {1.2 Gbps} from 5 to 1, {45 Mbps} from 2 to 1, and {25 Mbps} from 3 to 1. Figure \ref{fig:C613-Tput} shows the throughput performance of single-hop and multi-hop networks with route refinement and without route refinement. There are fluctuations in the throughput from 5 to 1 without route refinement after it switches to multi hop. We observe that the throughput is more stable with route refinement. It switch back to single hop after the blockage is removed. 
 \begin{figure*}[t]
 	\centering
 	 	\vspace{-.5cm}
 	\subfloat[\emph{Throughput performance} ]{  \includegraphics[scale=.3]{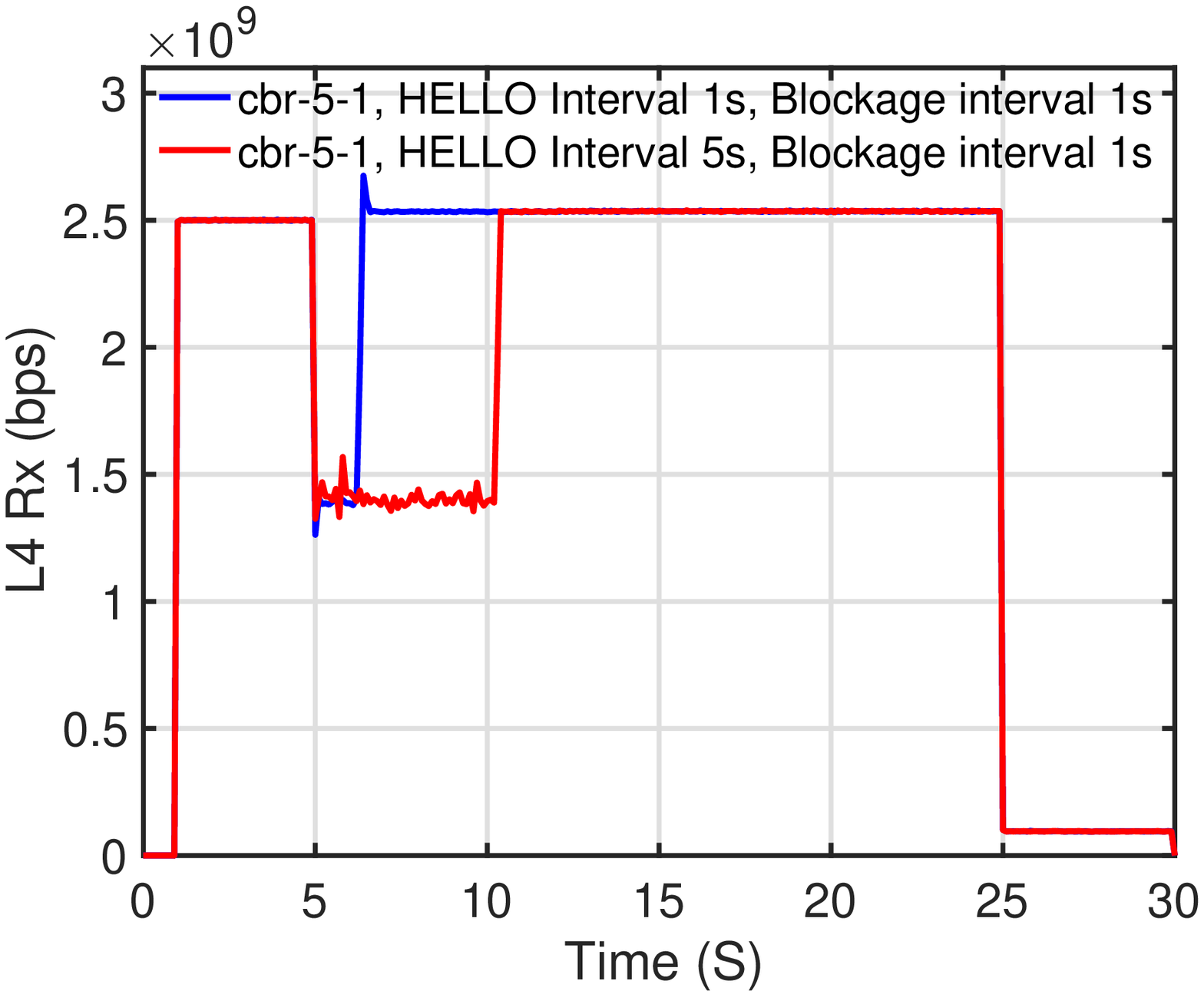}
 		\label{fig:B612-Tput}
 	}
 	\subfloat[\emph{Delay performance}]{
 		\includegraphics[scale=.3]{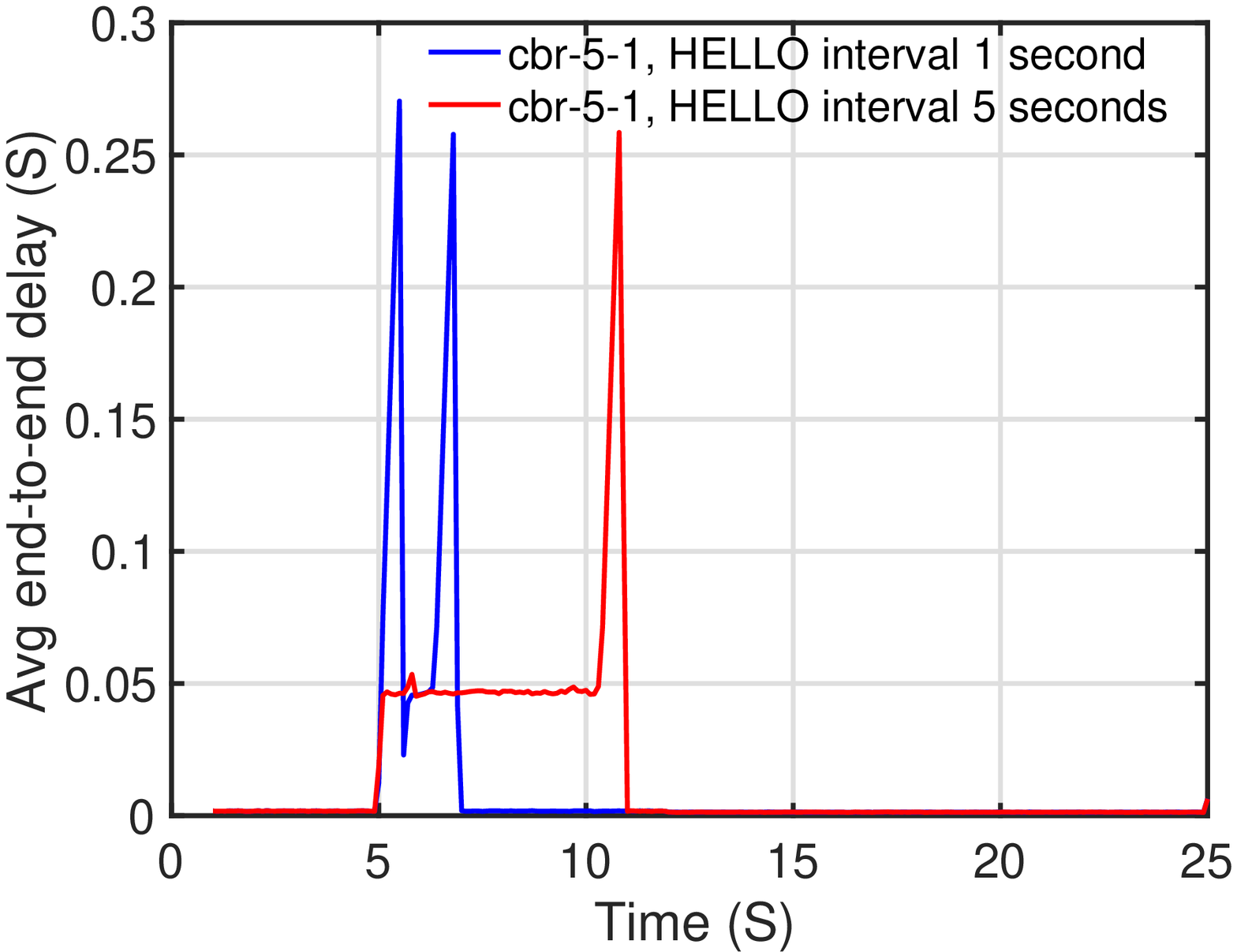}
 		\label{fig:B612-delay}
 	} 
 	\subfloat[\emph{CDF of delay}]{\includegraphics[scale=.3]{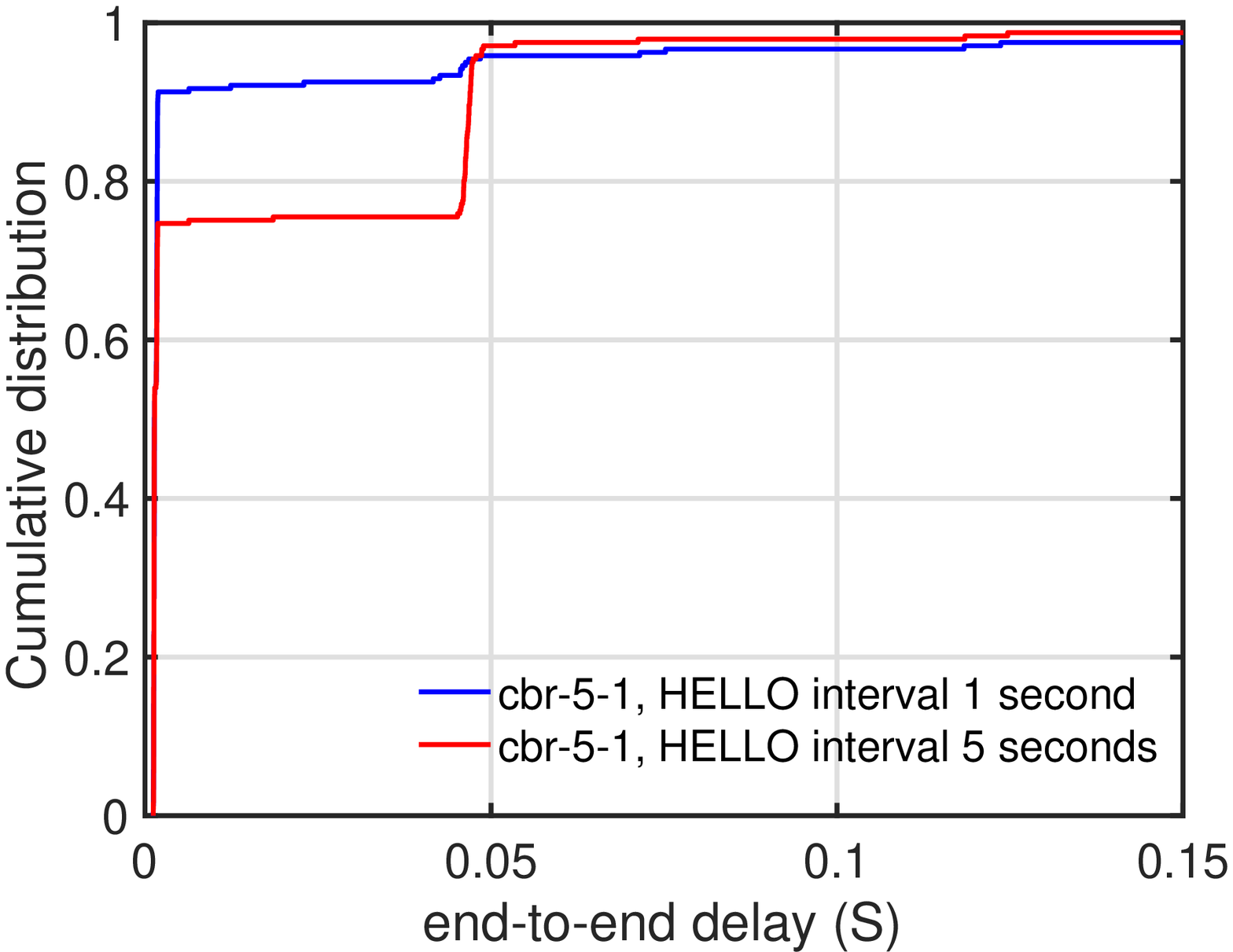}
 		\label{fig:B612-cdf-delay}
 	}
\vspace{-.2cm}
 	\caption[=]{\small{Performance of \textbf{Backpressure routing protocol} with different HELLO intervals. Blockage lasts for 1s. }}
 	\label{fig:hello-bcp}
 \end{figure*}
 \begin{figure*}[t]
 	\centering
 	 	\vspace{-.5cm}
 	\subfloat[\emph{Throughput performance} ]{  \includegraphics[scale=.3]{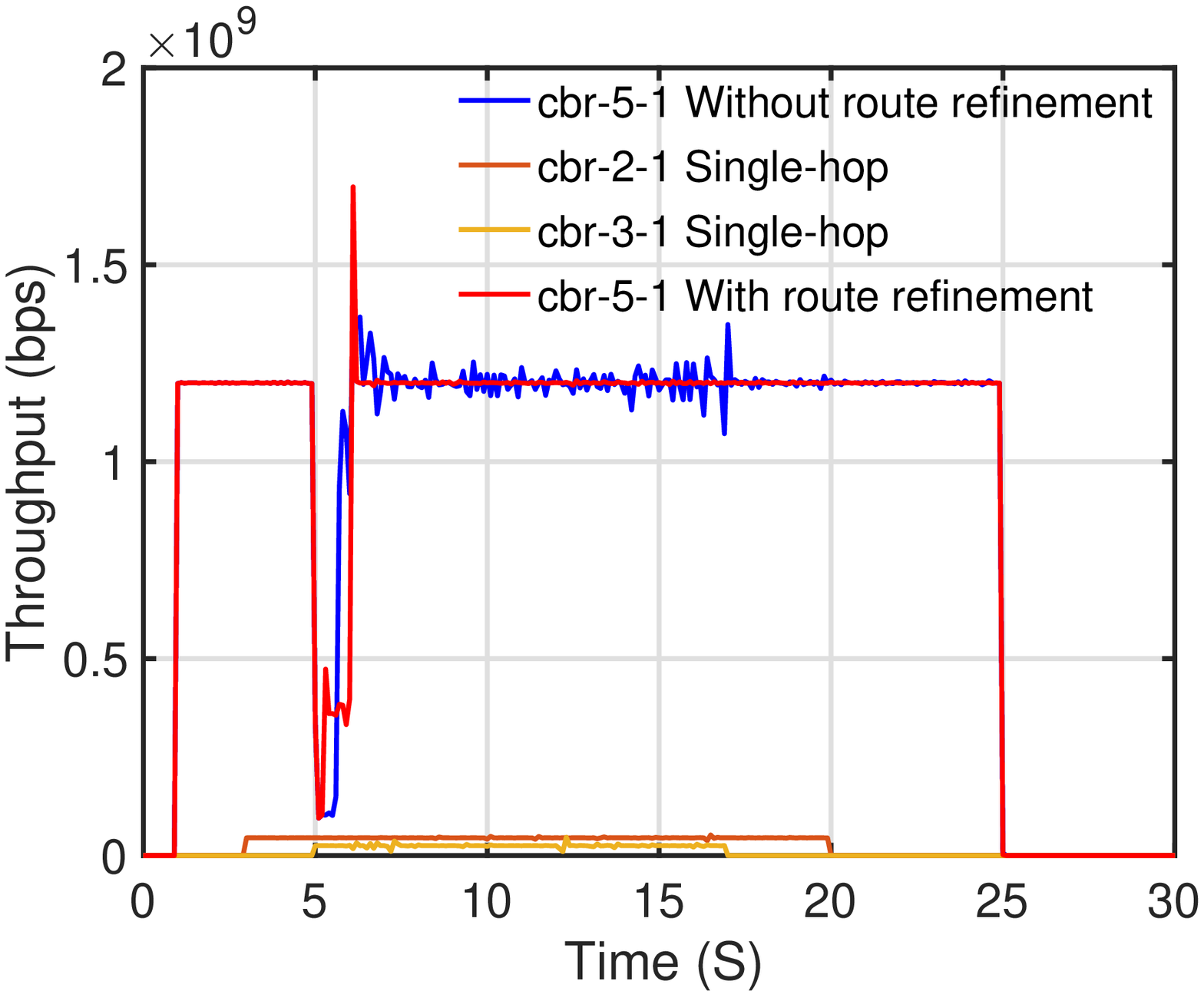}
 		\label{fig:C613-Tput}
 	}
 	\subfloat[\emph{Delay performance}]{
 		\includegraphics[scale=.3]{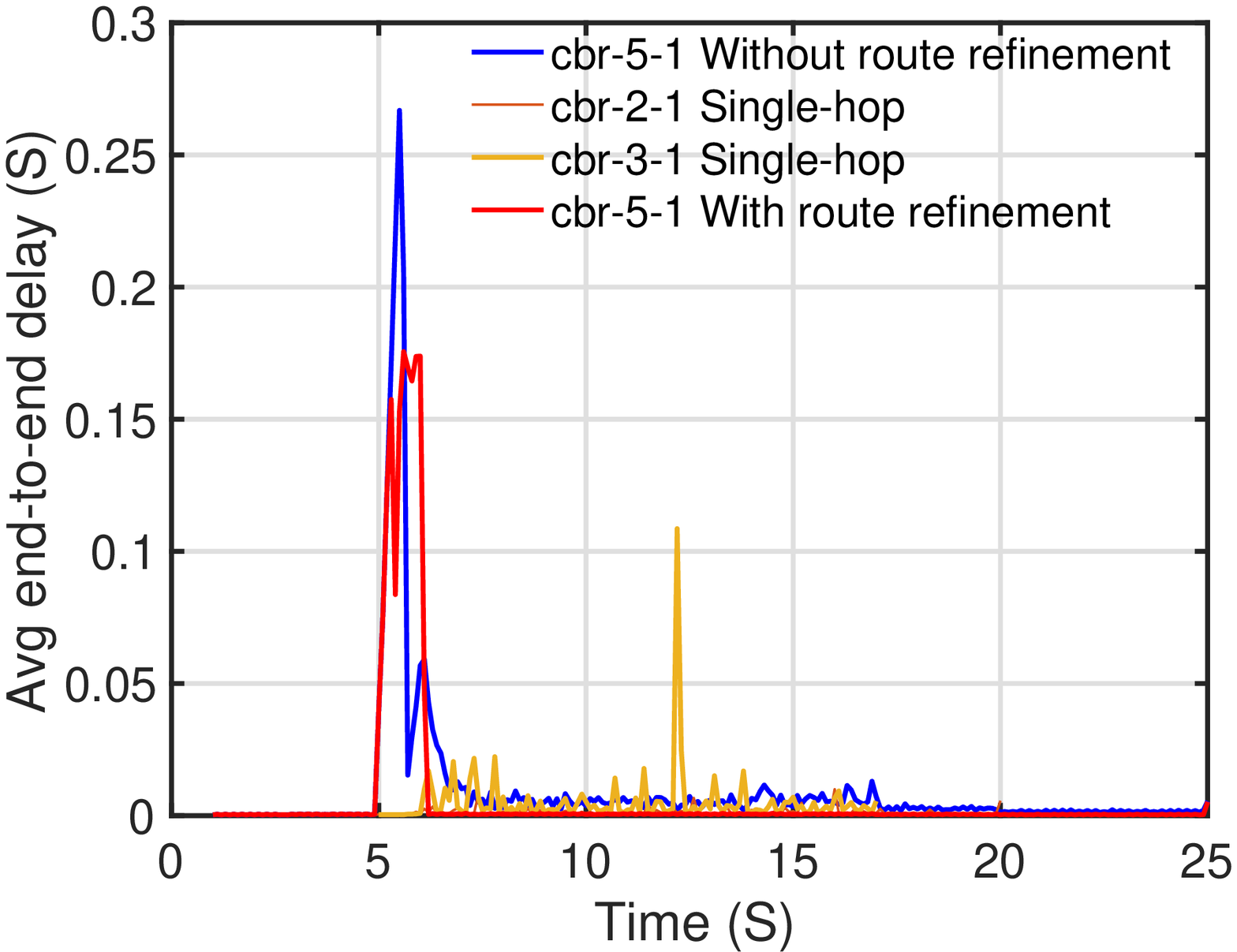}
 		\label{fig:C613-delay}
 	} 
 	\subfloat[\emph{CDF of Delay}]{\includegraphics[scale=.3]{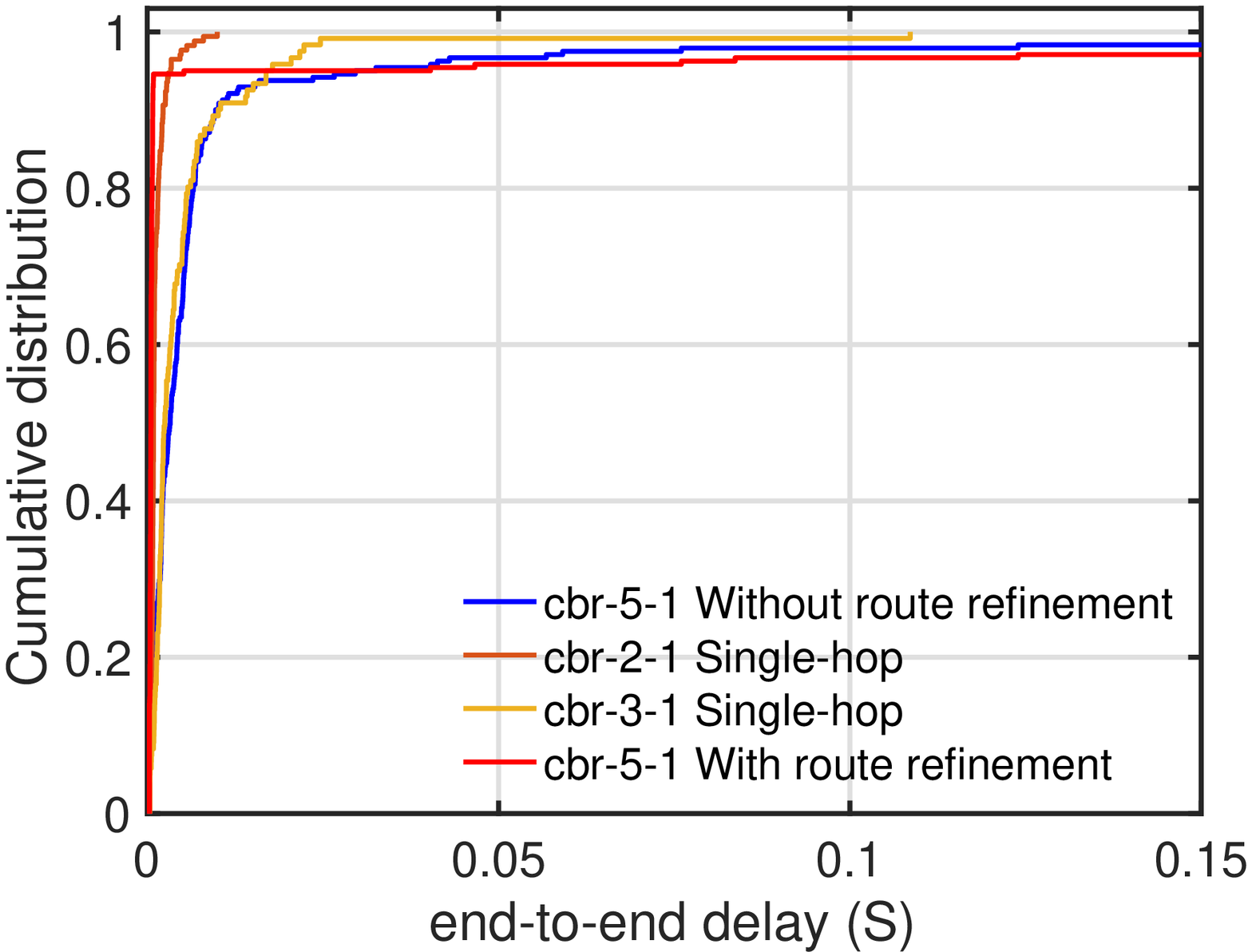}
 		\label{fig:C613-cdf-delay}
 	}
\vspace{-.2cm}
 	\caption[=]{\small{\textbf{AODV protocol with multiple data flows}, and with and without route refinement using SSW frames. Blockage lasts for 1s.}}
 \end{figure*}
In addition, Fig. \ref{fig:C613-cdf-delay}
compares the CDF of delay with route refinement and without route refinement. From the results, we observe that delay of the system with route refinement for CBR 5 to 1 is much smaller than the delay of the system without route refinement.}

\vspace{-.5cm}
\section{Conclusion}
\label{sec:conclusion}
In this paper, we proposed proactive route refinement schemes for AODV and Backpressure routing protocols. For the AODV protocol, we utilize sector sweep frames to transmit route request and route reply fields at each SSW interval. As a result,  multi-hop routes that have already been established by the on-demand routing protocol, are proactively refined using sector sweep frames in order to find more optimized routes as blockage dynamically changes. Our simulation results demonstrate that such a cross-layer protocol enhances the delay and throughput performance compared with when the route refinement is not activated.  For the Backpressure protocol, we proposed adding a periodic HELLO messages to distribute the necessary information needed to compute backpressure weights. Throughput and delay simulation results clearly demonstrate the role of the HELLO interval to achieve agile route refinement in Backpressure protocol. 

 \vspace{-.25cm}
\bibliographystyle{IEEEtran}
\bibliography{Ref-Hashemi}

\end{document}